\documentclass[11pt,superscriptaddress]{article}

\usepackage[a4paper, total={5.8in, 9in}]{geometry}
\usepackage[indent,skip=4pt]{parskip}
\usepackage{amsmath,amssymb,amsfonts,amsthm,mathrsfs,mathtools,cases}
\usepackage{dsfont}
\usepackage{authblk}
\usepackage[colorlinks=true,linkcolor=blue, citecolor=blue, bookmarks]{hyperref}
\usepackage{cleveref}
\usepackage{microtype}
\usepackage{comment}
\usepackage{graphicx,adjustbox} 
\usepackage{subfig}
\usepackage{wrapfig}
\usepackage[font=small,labelfont=bf]{caption}
\usepackage[dvipsnames]{xcolor}  
\usepackage{booktabs}
\usepackage{braket}
\usepackage[nottoc]{tocbibind}
\usepackage{nicematrix}
\usepackage{tikz}
\usepackage{pgfplots}
\usepackage{cite}
\pgfplotsset{compat=1.18}
\usepackage{enumitem}

\usetikzlibrary{matrix}
\usetikzlibrary{decorations.markings}
\usetikzlibrary{patterns}
\usetikzlibrary{arrows.meta}

\numberwithin{equation}{section}

\crefname{figure}{Fig.}{Figs.}
\crefname{equation}{Eq.}{Eqs.}
\crefname{section}{Sec.}{Secs.}
\crefname{appendix}{Appendix}{Appendices}
  
\colorlet{darkerblue}{MidnightBlue!20!black}
\colorlet{lightblue}{blue!70!white}

\hypersetup{
    colorlinks,
    linkcolor={Blue},
    citecolor={Blue},
    urlcolor={Blue},
}

\newcommand{\beq}{\begin{equation}}
\newcommand{\eeq}{\end{equation}}
\newcommand{\beqa}{\begin{eqnarray}}
\newcommand{\eeqa}{\end{eqnarray}}

\newcommand{\Tr}{\operatorname{Tr}}

\usepackage[framemethod=TikZ]{mdframed}

\newcommand{\abs}[1]{\left | #1 \right |}

\title{\textbf{Entanglement Hamiltonians in dissipative free fermions and the time-dependent GGE}}

\author[1]{Riccardo Travaglino}
\author[2]{Federico Rottoli}
\author[1]{Pasquale Calabrese}

\affil[1]{\textit{SISSA and INFN Sezione di Trieste, via Bonomea 265, I-34136 Trieste, Italy}}
\affil[2]{\textit{Dipartimento di Fisica dell’Universit\`a di Pisa and INFN Sezione di Pisa, Largo B. Pontecorvo 3, I-56127 Pisa, Italy.}}

\date{}

\begin{document}
\maketitle

\begin{abstract}
We investigate the dynamics of Entanglement Hamiltonians (EHs) in dissipative free-fermionic systems using a recent operator-based formulation of the quasiparticle picture. Focusing on gain and loss dissipation, we study the post-quench evolution and derive explicit expressions for the EH at the ballistic scale. In the long-time and weak-dissipation regime, the EH is shown to take the form of a time-dependent Generalized Gibbs Ensemble (t-GGE), with a structure that is universal across different initial states of the quench protocol. Within this framework, the emergence of the t-GGE is fully accounted for by the quasiparticle picture, and we argue that this description remains valid whenever the Lindbladian admits an appropriate coarse-grained representation.
\end{abstract}
\newpage

\tableofcontents

\newpage

\section{Introduction}
The relaxation of closed quantum systems toward locally stationary states has been the subject of intense investigation in the last two decades.
Motivated by key insight arising from the saturation of entanglement entropy in quantum quench protocols~\cite{
quench2,alba1,alba2,kimhuse,nahum,calabreserev}, substantial effort has been devoted to understanding how unitary dynamics can give rise to effective statistical ensembles in the long-time limit.  
It is now well established that, for generic (chaotic) systems, unitary time evolution leads to the emergence of thermal behaviour~\cite{PhysRevA.43.2046,srednicki1,Deutsch_2018,Rigol:2007juv,RevModPhys.83.863,DAlessio:2015qtq,Gogolin_2016} at late times when attention is restricted to subsystems that are sufficiently small compared to the full system. 
In contrast, for integrable systems, the relaxation leads to Generalized Gibbs Ensembles (GGEs), which explicitly incorporate the extensive set of (local) conserved quantities~\cite{Rigol,Fioretto_2010,VidmarRigol,fagessler,ilievski,calabrese2016introduction, essler2016quench,PIROLI2017362}. 
These novel perspectives on thermalisation as a key aspect of quantum evolution have led to significant advances in the understanding of transport and hydrodynamics in one-dimensional systems~\cite{DoyonOutOfEqCFT,NESS,NESS2,GHD1,GHD2,bastianello2022introduction,doyon2020lecture,ESSLERGHD}. Building on these ideas, it was subsequently recognized that hydrodynamics also provides an ideal framework for studying quench dynamics
\cite{Bertini_2018,Entanglement_GHD_interacting,alba2021generalized}. 
In particular, Refs.~\cite{rottoli2024entanglementhamiltoniansquasiparticlepicture,travaglino2024quasiparticlepictureentanglementhamiltonians} developed a hydrodynamic framework to study the post-quench evolution of the entanglement Hamiltonian (EH), defined as the logarithm of the reduced density matrix:
\begin{equation}
    \rho_A = \frac{e^{-K_A}}{Z_A}, \hspace{0.5cm} \rho_A = \Tr_{\overline{A}}\rho.
\end{equation}
Being constructed directly from $\rho_A$, the EH not only encodes  information about the entanglement spectrum (thus allowing one to reconstruct the Rènyi entropies, arguably the most relevant measures of entanglement in closed many-body systems~\cite{vidal_2003,Pasquale_Calabrese_2004,damico2008,Calabrese_2009}), but also contains information about the full structure of the eigenvectors of the reduced density matrix. 
It therefore provides the most complete characterization of the correlations within the system.
Consequently, it is rather surprising that closed-form expressions for the entanglement Hamiltonian can be obtained, primarily in quantum field theory~\cite{Bisognano:1975ih,Bisognano:1976za,Casini_2011,Cardy_2016,EH} and in free systems~\cite{ehFF1,EHFF2,Wen_2018,Eisler_2019,Di_Giulio_2020,Eisler_2020}. It is even more remarkable that, in these cases, the resulting entanglement Hamiltonian always exhibits some degree of locality.

The hydrodynamic approach developed in Ref.~\cite{rottoli2024entanglementhamiltoniansquasiparticlepicture} can be viewed as an operator-based reformulation of the well-known quasiparticle picture~\cite{quench1,quench2}. In this long-standing approach, quantum quenches at the ballistic scale generate pairs\footnote{In some contexts, a quench can generate multiplets of three or more entangled quasiparticles. While this modifies certain details of the ensuing dynamics, the core ideas of the approach can be generalized in a straightforward manner~\cite{bertini2018entanglement,bastianello_pair}.} of entangled quasiparticles, originating from the same point in space, which then propagate ballistically throughout the system, thereby spreading correlations.
Within this picture, the leading contribution to the time dependence of the entanglement entropy of a subsystem $A$ is determined by quasiparticle transport, since $S_A$ depends only on those pairs that are shared between the subsystem and its complement at a given time. 
The quasiparticle picture has been successfully applied to the study of a wide range of entanglement measures and characterizations in both free and interacting integrable systems~\cite{quench2,fagotti2008evolution, coser2014entanglement,alba1,alba2,groha2018full,alba2019quantum ,PIROLI2017362, alba2019scrambling,parez2021quasiparticle,Parez_2022,PRX,murciano2022negativity,murciano2022post,turkeshi2022,ares2023entanglement,ares2023lack,shion2,chalas2024quench,travaglino2025,Travaglino_2025_measurements}. In the latter case, its validity has been confirmed within the recently introduced framework of spacetime duality that reproduces the quasi-particle formulas for single-replica quantities, such as the von Neumann entanglement entropy~\cite{PRX,klobas2021,bertini_negativity,bertini2023nonequilibrium,bertini_asymmetric_2024,bertini_review}.

For open quantum systems, the dynamics is non-unitary, and the dissipative interaction between the system and its environment can often be modeled by a Lindblad evolution. Under these circumstances, the evolution of entanglement entropy is generally far more complicated. Even in free-fermionic systems, most Lindbladians of physical interest generate dynamics that do not preserve Gaussianity, making analytical progress difficult.
In the context of integrable systems, the effects of dissipation have nonetheless been studied extensively, largely motivated by experimental realizations. In particular, it has been shown that, for weak dissipation, Lindblad dynamics leads to relaxation toward generalized Gibbs ensembles with time-dependent chemical potentials. In this regime, dissipation manifests itself by inducing a slow dynamics at long time scales between such effective stationary states~\cite{Lange2017,tGGE}. 
This phenomenon, known as the time-dependent GGE (or t-GGE for short), has since been characterized in several theoretical studies and has also been explored experimentally~\cite{bouchoule2020,tGGE2,t_GGE_exp,tGGE3,lenarcic2024,lumia2025,PhysRevLett.129.220602,PhysRevResearch.6.013016}.
Moreover, in the context of free-fermion systems it has been shown that, when the dissipation preserves the Gaussianity of the time-dependent state, the quench dynamics of Rényi entropies and other entanglement measures can still be described by an appropriate modification of the quasiparticle picture~\cite{alba_carollo2021,carollo_alba2022,carollo2022,alba_carollo2023,Alba_2022,Caceffo_2024,caceffo2024fateentanglementquadraticmarkovian}. 
In the unitary case, results for various entanglement measures obtained from the quasiparticle picture can be traced back to a more fundamental quasiparticle description at the level of state evolution, which also explains the emergence of GGE at long times. It is therefore natural to expect that an analogous quasiparticle picture can be formulated in the presence of dissipation as well.
Such a picture should operate at the level of the evolving state, and hence of the entanglement Hamiltonian, thereby implying a corresponding structure for entanglement measures.
In this work we address precisely this problem by generalising the hydrodynamic framework of Refs.~\cite{rottoli2024entanglementhamiltoniansquasiparticlepicture,travaglino2024quasiparticlepictureentanglementhamiltonians} to account for the presence of dissipation. 
In particular, focusing on gain and loss dissipation, we show that the effective approach of Ref.~\cite{rottoli2024entanglementhamiltoniansquasiparticlepicture} can be modified consistently and straightforwardly, as the corresponding Lindbladian admits a convenient coarse-grained description. 

The structure of the paper is the following. In section~\ref{sec:setup} we introduce the Lindbladian and quench protocol of interest. In section~\ref{sec:review} we review the approach in the unitary case, and in~\ref{sec:coarsegraindiss} we discuss its extension to the dissipative case. Sections~\ref{sec:EH} and~\ref{sec:tGGE} are the core of the paper, in which we construct the EH and we show that taking the proper limits this reduces to a t-GGE. Finally, we numerically verify our expressions in section~\ref{sec:numerics} before concluding in~\ref{sec:concl}.

\section{Setup}
\label{sec:setup}
We consider a lattice system of spinless free fermions $\{c_i,c^\dagger_j\}=\delta_{ij}$, with time evolution determined by the one dimensional tight binding  Hamiltonian,
\begin{equation}
    H = -\frac{1}{2} \sum _{i=1}^L c_i^\dagger c_{i+1} + h.c. = -\sum_{k}\cos  k \, c_k^\dagger c_k.
\end{equation}
with the quasiparticle velocity being $\varepsilon'_k=\sin k$. We consider a quench protocol, namely the many body evolution originating from a specific initial state $\ket{\psi_0}$ which in general has a complicated representation in terms of the eigenstates of the Hamiltonian. Some typical choices include  coherent (or squeezed) states
\begin{equation}
    \ket{\psi_0} \propto \exp\!\left\{\sum_{x,y} \mathcal{M}(x-y)c_x^\dagger c_y^\dagger\right\} \ket{0},
    \label{eq:coherent}
\end{equation}
where $\mathcal{M}(x-y)$ is an antisymmetric function. This BCS-like state is ubiquitous in the context of free fermionic quenches because of its relation with Bogoliubov transformations~\cite{fagotti2008evolution,Fioretto_2010}, and its relations to boundary integrability~\cite{Ghoshal:1993tm,PIROLI2017362}. A second special class of initial states sharing the same integrable feature are characterized by two-site shift invariance, 
\begin{equation}\label{eq:symm_preserve_state_1d}
    \ket{\psi}=\prod_k \left (\sqrt{n_k} c^\dagger_k +e^{i\varphi_k}\sqrt{1-n_k} c^\dagger_{k-\pi} \right )\ket{0}.
\end{equation}

In previous works~\cite{rottoli2024entanglementhamiltoniansquasiparticlepicture, travaglino2024quasiparticlepictureentanglementhamiltonians, travaglino2025}, the evolution of entanglement and negativity Hamiltonians following quenches from these two types of initial states has been studied using a coarse-graining procedure, which will be reviewed in the next section. In this work, we extend the discussion to dissipative systems undergoing non-unitary dynamics, 
focusing on a specific form of Lindbladian describing gain and loss dissipation,
\beq
    \mathcal{L}[\rho] = -i[H,\rho] + \gamma_L\sum_i \left(c_i \rho c_i^\dagger - \frac{1}{2}\{\hat{n}_i,\rho\}\right)
    +\gamma_R\sum_i \left(c_i^\dagger \rho c_i - \frac{1}{2}\{1-\hat{n}_i,\rho\}\right),
    \label{eq:gainlossoriginal}
\eeq
where $\hat{n}_i = c_i^\dagger c_i$ is the number operator at site $i$.
The interest in Lindbladians of this form is twofold: first of all they model the loss or gain of particles that can take place when a physical system is coupled to an environment; secondly, it is analytically tractable as it preserves the Gaussianity of the state $\rho$. 
In fact, it is known that in fermionic systems, Lindblad operators which are linear in the fermionic operators generate Gaussianity preserving channels~\cite{Barthel_2022,Prosen_third_quantization}.
Therefore, at all times during the evolution, all correlation functions are fully determined by the correlation matrix via Wick’s theorem. 
The correlation matrix is
\begin{equation}
    C_{x,y}(t) = \braket{\boldsymbol c_x^\dagger \boldsymbol c_y}(t), \hspace{0.4cm} \boldsymbol c^\dagger_x = \begin{pmatrix}
    c_x \\ c_x^\dagger  \end{pmatrix},
\end{equation}
where for states which commute with the total number operator $\hat{N} = \sum_x c_x^\dagger c_x$ it is sufficient to consider the restriction to the particle number preserving component $\braket{c_x^\dagger c_y}$. 
In particular, under non-unitary dynamics with Lindbladian~\eqref{eq:gainlossoriginal}, the correlation matrix evolves as
\begin{equation}
    C_{xy}(t) =   e^{-\Gamma t} C_{xy}^{\rm u}(t) + \overline{n}(1-e^{-\Gamma t}) \delta_{x,y} \label{eq:correlationgainloss}
\end{equation}
where $C_{xy}^{\rm u}(t)$ is the unitary correlation matrix which is obtained through the pure Hamiltonian evolution, and we have parametrized the solution through the total rate $\Gamma = \gamma_L + \gamma_G$ and the asymptotic density $\overline{n}= \gamma_G/\Gamma$.
For symmetric initial states, in which the anomalous correlations $\left < c_x c_y \right >$ vanish, the single particle entanglement Hamiltonian $k_A$ is related to the correlation matrix through Peschel's formula~\cite{PhysRevB.69.075111,Peschel_2009}
\begin{equation}
    C_A(t) = \frac{1}{1+e^{k_A{(t)}}},\label{eq:peschelformula}
\end{equation}
where $C_A(t)$ is the restriction of the correlation matrix to the subsystem of interest, and $k_A{(t)}$ is the matrix of coefficients of the entanglement Hamiltonian expanded in the natural fermionic basis,
\begin{equation}\label{eq:peschel}
K_A(t)= \sum_{\boldsymbol{i},\boldsymbol{j}}k_{A,\boldsymbol{i},\boldsymbol{j}}{(t)}c_{\boldsymbol{i}}^\dagger c_{\boldsymbol{j}}.
\end{equation}
While~\eqref{eq:peschelformula} allows us to obtain the exact EH, it is not transparent and provides no insight on the structure of the solution. In this work we will use this expressions to validate numerically the results in the ballistic regime obtained through the quasiparticle picture.

\section{Review of the hydrodynamic picture}
\label{sec:review}
In this section we show how to implement a coarse graining of the density matrix in the simplest case which was studied in Ref.~\cite{rottoli2024entanglementhamiltoniansquasiparticlepicture}.
Let us assume that the initial state is a squezeed state of the form
\beq
    \ket{\psi} =  \frac{1}{\mathcal{N}}\,\exp\!{\left (\sum_{x,x'=1}^L \mathcal{M}(x-x')c^\dagger_{x}c^\dagger_{x'}\right )}\ket{0},
\eeq
with $\mathcal{M}(x)$ an odd function of $x$. 
The coarse-graining is implemented by dividing the system into fluid cells of size $\Delta$, where we assume separation of scales $a \ll \Delta \ll \ell$, with $a$ the lattice spacing and $\ell$ the size of the system.
The lattice coordinate naturally splits into two parts, $x=x_0+z$,  where $x_0\in [1,L/\Delta]$ indexes the fluid cell to which  $x$ belongs and $z\in[0,\Delta-1]$ gives the microscopic coordinate inside the fluid cell.
If the correlations of the initial state decay fast enough compared to the size of the fluid cells, we can truncate the coefficients as
\begin{equation}
    \label{eq:fluid_cell}\mathcal{M}(x - x')\approx 0, \qquad\forall \abs{x -x' } > \xi,\,\xi\ll\Delta ,
\end{equation}
 thus allowing to simplify the initial state to
\begin{align}
    \ket{\psi} &= \frac{1}{\mathcal{N}}\,\exp\!{\left (\sum_{x_0=1}^{L/\Delta}\sum_{z,z'=0}^{\Delta-1} \tilde{\mathcal{M}}(z-z')c^\dagger_{x_0+z}c^\dagger_{x_0+z'}\right )}\ket{0}\\\label{eq:fluid_initial_state}
    &=\frac{1}{\mathcal{N}}\prod_{x_0=1}^{L/\Delta} \exp\!{\left (\sum_k\mathcal{M}(k)b^\dagger_{x_0,k}b^\dagger_{x_0,-k}\right )}\ket{0}
\end{align}
where we have introduced the coarse-grained creation and annihilationoperators $b_{x_0,k}^\dagger$, $b_{x_0,k}$, defined by taking the Fourier transform only inside the fluid cell
\begin{equation}
    b^\dagger_{x_0,k}=\frac{1}{\sqrt{\Delta}} \sum_{z=0}^{\Delta-1} e^{-ikz} c^\dagger_{x_0+z}, \quad b_{x_0,k} =\frac{1}{\sqrt{\Delta}} \sum_{z=0}^{\Delta-1} e^{ikz} c_{x_0+z}.
    \label{eq:fluidcelloperators}
\end{equation}
The fluid cell momenta $k$ appearing in the definition~\eqref{eq:fluidcelloperators} run over a set of values quantized according the fluid cell size $\Delta$, i.e. $k\in \frac{2\pi}{\Delta}\mathbb{Z}_\Delta$.

Expression~\eqref{eq:fluid_initial_state} allows us to express the initial state in a simplified fashion as
\begin{equation}
\label{eq:product}
    \rho{(0)} = \bigotimes_{x_0}\bigotimes_{k>0} \rho_{x_0,k},
\end{equation}
where $\rho_{x_0,k}$ is the density matrix of a pair of quasiparticles of opposite momenta inside  the cell located at $x_0$. Given the pair structure, the product is restricted to  $k>0$ to avoid over counting. Representing the density matrix of the post-quench vacuum state as
\begin{equation}
\ket{0}\bra{0}=\bigotimes_{x_0}\bigotimes_{k}(1-\hat{n}_{x_0}(k)),
\end{equation}
where we have introduced the coarse-grained particle number operator $\hat{n}_{x_0}(k)=b^\dag_{x_0,k}b_{x_0,k}$, we find the pair density matrix
\begin{equation}
\begin{split}
    \rho_{x_0,k} =&\, 
    n_k \hat{n}_{x_0}(k)\hat{n}_{x_0}(-k)   + (1-n_k)(1- \hat{n}_{x_0}(k))(1- \hat{n}_{x_0}(-k)) \\ &+  \sqrt{n_k(1-n_k)}(e^{i\varphi_k}b_{x_0,k}^\dagger b_{x_0,-k}^\dagger +e^{-i\varphi_k}  b_{x_0,-k}b_{x_0,k}).
    \label{eq:densitymatrix}
\end{split}
\end{equation}
As shown in Ref.~\cite{travaglino2024quasiparticlepictureentanglementhamiltonians}, both the occupation functions $n_k$ and the phase $e^{i\varphi_k}$ can be expressed in terms of the coefficients $\mathcal{M}(k)$.
Note that the occupation functions $n_k$ do not acquire a spatial dependence, even if the operators $\hat{n}_{x,k}$ do, owing to the homogeneity of the quench.
For the symmetric states~\eqref{eq:symm_preserve_state_1d} the discussion is similar, with the caveat that the density matrix represents particle-hole entangled pairs
\begin{equation}\begin{split}
    \rho_{x_0,k}{(t)} =&\, n_k\hat{n}_{x_0}(k)(1-\hat{n}_{x_0}(k-\pi)) + (1-n_k)(1-\hat{n}_{x_0}(k)) \hat{n}_{x_0}(k-\pi) 
    \\
    &+ \sqrt{n_k(1-n_k) }\left(e^{i\varphi_k}b_{x_0(k),k}^\dagger b_{x_0,k-\pi}+e^{-i\varphi_k}b_{x_0,k-\pi}^\dagger b_{x_0(k),k}\right).
    \label{eq:symm_presrve_pure} 
\end{split}\end{equation}
Some typical realizations of this kind of states are the Néel state, with constant occupation functions $n_k=1/2$, and the dimer state, for which $n_k= \left (1\pm \cos k \right ) /2$ (the sign depending on the convention). In this work, we will focus mostly on this class of states, and comment on the symmetry breaking case~\eqref{eq:coherent} in the appendix.

Once the initial state has been expressed in terms of coarse grained operators, a second approximation  regards the evolution of the operators; as shown in appendix~\ref{App}, at leading order we have\footnote{Note that the phase appearing here was not considered in previous works within this approach, as the quantities of interest typically depended only on combinations of the form $b^\dag_{x,k} b_{x,k}$, for which the phase cancels out. In the present context, however, this phase plays a nontrivial role as a consequence of dissipation.}
\begin{equation}\label{eq:QPP_heisenber_evolution}
e^{iHt}b^\dag_{x_0, k} e^{-iHt} \approx e^{it\varepsilon_k}b^\dag_{x_0+ \varepsilon'_kt, k}. 
    \end{equation}
According to \cref{eq:QPP_heisenber_evolution}, the quasiparticles are  coherently transported between fluid cells according to their velocity $\varepsilon'_k$, starting at $t = 0$ from the cell indexed by $x_0$ and reaching at time $t$ the one indexed by $x_t(k)=x_0+\varepsilon'_k t$. Any dispersive effects are assumed to be subleading on length scales comparable to $\Delta$, and the main quantum feature of the evolution is given by the presence of the phase $e^{it\varepsilon_k}$ which describes the internal evolution of the degrees of freedom of the wavepacket. 

Once the evolution of the operators is known, the time evolved density matrix for a single pair originating from $x_0$ can be easily found, giving in the symmetry breaking case~\eqref{eq:densitymatrix}
\begin{equation}\begin{split}
    \rho_{x_0,k}(t) =&\, 
    n_k \hat{n}_{x_t(k)}\hat{n}_{x_t(-k)}    + (1-n_k)(1- \hat{n}_{x_t(k)})(1-\hat{n}_{x_t(-k)}) \\ 
    &+  \sqrt{n_k(1-n_k)}(e^{i\varphi_k(t)}b_{x_t(k),k}^\dagger b_{x_t(-k),-k}^\dagger +e^{-i\varphi_k(t)} b_{x_t(-k),-k}b_{x_t(k),k}).
    \label{eq:densitymatrixt}
\end{split}\end{equation}
where $ \hat{n}_{x_t(\pm k)}=b^\dag_{x_t(\pm k),\pm k}b_{x_t(\pm k),\pm k}$, and $\varphi_k(t) = \varphi_k + 2t\varepsilon_k$. For simplicity, we will refer to the right mover in the pair through $x_t^+$ and to the left mover as $x_t^-$ and leave the $k$ dependence  implicit. In this notation the evolution in the symmetric case is 
\begin{equation}\begin{split}
    \rho_{x_0,k}{(t)} =&\, n_k\hat{n}_{x_t^+}(1-\hat{n}_{x_t^-}) + (1-n_k)(1-\hat{n}_{x_t^+} )\hat{n}_{x_t^-} \\\label{eq:evolved_symmetrypreserved} &+ \sqrt{n_k(1-n_k) }\left(e^{i\varphi_k(t)}b_{x_t^+}^\dagger b_{x_t^-}+e^{-i\varphi_k(t)}b_{x_t^-}^\dagger b_{x_t^+}\right).
\end{split}\end{equation}
The construction of the coarse-grained density matrix is now complete, allowing us to write explicitly the reduced density matrix and consequently the entanglement Hamiltonian by exponentiation. 
In fact, for a subsystem $A$, the reduced density matrix is given by the configurations in which the right mover is in $A$ and the left mover is in $\overline{A}$ (or viceversa), i.e. 
\begin{equation}\begin{split}\label{eq:formweneed}
    \Tr_{\overline{A}}[\rho_{x_0,k}(t)] &= n_k \hat{n}_{x_t^+} + (1-n_k)(1-\hat{n}_{x_t^+})\\
    &= \frac{1}{1+e^{-\eta(k)}}\exp\!{\left(-\eta(k)\hat{n}_{x_t^+}\right)}
\end{split}\end{equation}
where  in the second line we have introduced 
\begin{equation}\label{eq:eta}
    \eta(k) = \log\left(\frac{1-n_k}{n_k}\right).
\end{equation}
Note that the phase factor $e^{i\varphi_k(t)}$ drops out, as it  appears only in the off-diagonal terms, leaving only the dependence on $n_k$.   
Summing over all modes and fluid cells we determine that $\rho_A(t)$ takes the following form
\begin{equation}\label{eq:decomposition}
    \rho_A(t) \approx \rho_{\rm mixed}(t)\otimes \rho_{\text{pure}}{(t)},
\end{equation}
which is split into a mixed state part containing the entanglement Hamiltonian and a pure part arising from the case when both particles in the pair are in $A$\footnote{One might wonder why it is justified to retain only the component $\rho_\text{mixed}$ in the definition of the entanglement Hamiltonian. The reason is that, by construction, $\rho_\text{pure}$ represents a pure state and is therefore a projector.  As such, its logarithm is ill-defined: since its spectrum consists only of eigenvalues equal to $0$ and $1$, even restricting the logarithm to the subspace of nonzero eigenvalues would yield a vanishing contribution to the  EH.}. The mixed part is then 
\begin{align}
    \rho_{\rm mixed}(t) &= \frac{1}{
\mathcal{Z}_A}\,e^{-K_{A}(t)},\\\label{eq:ent_ham_mixed}
    K_{A}(t)&=\sum_{k}\sum_{\{x_0|x_{t}(k)\in A\,\& \, x_{t}(-k)\notin A\}}\eta(k)\,\hat{n}_{x_t}(k),
\end{align}
where $\mathcal{Z}_A=\operatorname{Tr}[e^{-K_{A}}]$. Expressing the EH in terms of the coarse grained operators allows us to immediately evaluate the Rènyi entropies, which reproduce correctly the known results for the extensive part of the entropy.
To unveil the correlation structure, we transform back to real space, 
\begin{equation}
   K_{A}(t) = \int dx \int dz \,\mathcal{K} (x,t,z)\, c_{x}^\dag c_{x+z},
   \label{eq:realspaceunitaryK}
\end{equation}
where the kernel $\mathcal{K} (x,t,z)$ takes the form
\begin{equation}
    \mathcal{K} (x,t,z) = \int \frac{dk}{2\pi}\, \chi_{A\overline{A}}(x,k,t)\, \eta(k)\, e^{-ikx} ,
\end{equation}
and $\chi_{A\overline{A}}(x,k,t) $ is a counting function which selects shared pairs; 
this is evaluated in Ref.~\cite{rottoli2024entanglementhamiltoniansquasiparticlepicture}, and it satisfies $\int dx\, \chi_{A\overline{A}}(x,k,t)  = \min(2|\varepsilon'_k|t,\ell_A)$  which 
yields the correct result for the evaluation of the entropy.

\section{Coarse graining with dissipation}
\label{sec:coarsegraindiss}
In this section we include dissipative effects to the hydrodynamic framework introduced in the previous section. We start by considering the simplest possible Lindbladian evolution, in which we only act with Lindblad operators $L_i = c_i$, representing pure loss. The equation of motion of the density matrix is
\begin{equation}
    \frac{d}{dt}\rho =\mathcal{L}[\rho] = -i[H,\rho] + \gamma\sum_i \left(c_i \rho c_i^\dagger - \frac{1}{2}\{n_i,\rho\}\right).
    \label{eq:originallindblad}
\end{equation}
The crucial observation is that when we express the Lindblad equation~\eqref{eq:originallindblad} in terms of the coarse-grained operator, the form of the equation is preserved
\begin{equation}\begin{split}
    \mathcal{L}[\rho] &= -i[H,\rho] + \gamma\sum_{x,z} \sum_{k,k'}  e^{ -i z(k-k')}\left(b_{x,k} \rho \, b_{x,k'}^\dagger - \frac{1}{2}\{b_{x,k'}^\dagger b_{x,k},\rho\}\right)  \\
    &=  -i[H,\rho] + \gamma\sum_{x,k} \left(b_{x,k} \rho \, b_{x,k}^\dagger - \frac{1}{2}\{b_{x,k}^\dagger b_{x,k},\rho\}\right),
    \label{eq:simpleLindbladian}
\end{split}\end{equation}
where in the second line we have summed over $z$ to give a Kroenecker delta $\delta_{k,k'}$. Note that this is a non-trivial statement: it is easy to convince oneself that general non-Gaussian dissipations (such as simple dephasing in the $c^\dag_x c_x$ basis) do not have this property. 
In the presence of both gain and loss of particles,
the discussion can be generalized straighforwardly to obtain
\begin{equation}
\begin{split}
    \mathcal{L}[\rho]
    =\,  -i[H,\rho] &+ \gamma_L \sum_{x,k} \left(b_{x,k} \rho \, b_{x,k}^\dagger  - \frac{1}{2}\{b_{x,k}^\dagger b_{x,k},\rho\} \right) \\
    &+ \gamma_G \sum_{x,k} \left( b^\dagger_{x,k} \rho \, b_{x,k}- \frac{1}{2}\{ b_{x,k} b_{x,k}^\dagger,\rho\}\right).
    \label{eq:gainlossLindbladian}
\end{split}
    \end{equation}
Hence we see that the dissipative term has exactly the same structure when written in terms of the coarse grained operators. 
Since~\eqref{eq:gainlossLindbladian} has the same structure as~\eqref{eq:gainlossoriginal} with coarse grained operators instead of microscopic ones, the correlation matrix built from the former has the same structure
\begin{equation}
    \braket{b_{x,k}^\dag b_{y,q}}(t) =  e^{-\Gamma t}  \braket{b_{x,k}^\dag b_{y,q}}^{\rm u}(t) + \overline{n}(1-e^{-\Gamma t}) \delta_{x,y}\delta_{k,q}.
    \label{eq:correlationBoperators}
\end{equation}
Moreover, to evaluate higher correlators we apply Wick theorem also to correlators involving $b_{x,k}$ and $b_{x,k}^\dagger$, which, e.g., yields  the following expression for the four-point functions 
\begin{equation}
     \braket{\hat n_{x,k} \hat n_{y,q}}(t) = \braket{\hat n_{x,k}} \braket{ \hat n_{y,q}} - \braket{b_{x,k}^\dag b_{y,q}} \braket{b_{y,q}^\dag b_{x,k}}.
     \label{eq:wick}
\end{equation}

As reviewed in the previous section, a crucial aspect of the quasiparticle picture in the unitary case is the decomposition of the density matrix in a product form~\eqref{eq:product} which is then preserved by the evolution. 
In fact, this structure is also preserved by the addition of dissipation: given the result~\eqref{eq:correlationBoperators} and Wick theorem, the dissipation does not add correlations between pairs of quasiparticles which are not correlated already through the unitary evolution. 
This simple argument leads to the conclusion that also in the evolution generated by the Lindbladian~\eqref{eq:gainlossLindbladian} the state has at all times the form
\begin{equation}
    \rho = \bigotimes_{x,k} \rho_{x,k}(t),
\end{equation}
where the evolution of $\rho_{x,k}(t)$ in the dissipative regime exhibits non-trivial modifications compared to the unitary case~\eqref{eq:evolved_symmetrypreserved}, while still preserving the underlying quasiparticle-pair structure. 
The most general form of the time evolved pair is 
\begin{equation}\begin{split}
    \rho_{x,k}{(t)} =&\, \alpha \hat{n}_{x_t^+}\hat{n}_{x_t^-} + \beta \hat{n}_{x_t^+}(1-\hat{n}_{x_t^-} )+\gamma (1-\hat{n}_{x_t^+} ) \hat{n}_{x_t^-}\\\label{eq:evolved_symmetrypreserved_dissipation} &+  \delta (1-\hat{n}_{x_t^+} )(1-\hat{n}_{x_t^-} )  + \varepsilon\left(e^{i\varphi_k(t)}b_{x_t^+}^\dagger b_{x_t^-}+e^{-i\varphi_k(t)}b_{x_t^-}^\dagger b_{x_t^+}\right),    
\end{split}\end{equation}
where again we introduced $x_t^\pm = x_0 \pm \varepsilon'_k\, t$ to lighten the notation.
In order to determine the coefficients in \cref{eq:evolved_symmetrypreserved_dissipation}, we impose that the correlation functions computed from $\rho_{x,k}{(t)}$ are equal to those obtained from \cref{eq:correlationBoperators,eq:wick}.
This leads to the identification 
\begin{equation}
    \begin{cases}
        \alpha = \tilde{0}_\gamma \tilde{1}_\gamma,  \\
        \beta = n_ke^{-\Gamma t} + \tilde{0}_\gamma(1-\tilde{1}_\gamma), \\
        \gamma = n_k(\tilde{0}_\gamma - \tilde{1}_\gamma) + \tilde{1}_\gamma (1-\tilde{0}_\gamma), \\
        \delta = (1-\tilde{1}_\gamma)1-\tilde{0}_\gamma, \\
        \varepsilon = \sqrt{n_k(1-n_k)} (\tilde{1}_\gamma - \tilde{0}_\gamma),
    \end{cases}
    \label{eq:coefficients}
\end{equation}
where we have introduced the dissipation-induced dressing, which will be used in the following
\begin{equation}
     \tilde a_\gamma = e^{-\Gamma t}a +\overline{n} (1-e^{-\Gamma t}),\label{eq:tilded}
\end{equation}
which is implementing the transformation~\eqref{eq:correlationgainloss} to generic functions or numbers. 
This transformation describes the evolution under dissipation of the occupation functions, and therefore $\tilde{0}_{\gamma}$ and $\tilde{1}_\gamma$ represent the evolution of states with constant initial occupation functions $0$ and $1$ respectively, which yields
\begin{equation}
    \tilde{0}_\gamma = \overline{n}\left ( 1 - e^{-\Gamma t} \right ), \qquad \tilde{1}_\gamma = e^{-\Gamma t } + \overline{n}\left ( 1 - e^{-\Gamma t} \right ).
\end{equation}
Notice that the structure~\eqref{eq:evolved_symmetrypreserved_dissipation} implies $\Tr[\rho_{x,k}^2]\neq1$, as expected since the full state becomes mixed under the dissipative dynamics.
In particular, this means that, unlike in the unitary case, pairs for which both quasiparticles lie within  $A$ also contribute to the Rényi entropies, and hence to the entanglement Hamiltonian. 

\section{Entanglement Hamiltonian}
\label{sec:EH}
The final discussion of the previous section implies that the EH will consist of two main contributions: one containing pairs which are shared between $A$ and the complement, which we will refer to as the quantum contribution to match the nomenclature of~\cite{alba_carollo2021}, and the second one containing pairs in which both particles are in $A$, which represent a purely classical dissipation-induced contribution
\begin{equation}
    K(t) = K_{\rm q}(t)  + K_{\rm c}(t).
\end{equation}
In general, the number of pairs which contribute to the classical part vanishes at long times, as all pairs originated within the system are eventually transported out of it,  while the number of shared pairs saturates to the size of the subsystem.
This has the consequence that at large times the classical contribution disappears, while the asymptotic state is completely determined by the quantum part. 
Clearly, if $t$ is taken to infinity at finite $\Gamma$, the correlation matrix would become diagonal, as follows from \cref{eq:correlationBoperators}, thereby trivializing the dynamics.
Instead, by keeping $\Gamma t = \text{const} $ while taking the large time limit with $t$ much larger than the subsystem size $\ell_A$, we will show that the quantum contribution gives rise to a t-GGE.

\subsection{The quantum contribution}
Considering $\rho_{x,k}$, we focus on the situation in which the right mover is inside $A$ and the left mover is outside. Tracing over the complement $\overline{A}$, one obtains after some algebra
\begin{equation}
    \Tr_{\overline{A}}[\rho_{x,k}] = \hat{n}_{x_t^+} \left[e^{-\Gamma t}n +\overline{n} (1-e^{-\Gamma t})\right] + (1-\hat{n}_{x_t^+}) \left[ e^{-\Gamma t}(1-n)+1-e^{-\Gamma t} - \overline{n} (1-e^{-\Gamma t})\right]\!.
    \label{eq:quantum_trace_difficult}
\end{equation}
This expression can be simplified by writing it in terms of the dressed occupation function $\tilde{n}_\gamma$, defined according to \cref{eq:tilded}.
With this definition, \cref{eq:quantum_trace_difficult} reduces to a simple expression which can be immediately exponentiated,
\begin{equation}
     \Tr_{\overline{A}}[\rho_{x,k}] =\tilde n_\gamma\hat{n}_{x_t^+} + (1-\tilde n_\gamma) (1-\hat{n}_{x_t^+}) = (1-\tilde n_\gamma)\exp\left\{-\log\left(\frac{1-\tilde n_\gamma}{\tilde n_\gamma}\right)\hat{n}_{x_t^+}\right\}.
\end{equation}
Notice that this formula is a simple generalization of the result of Ref.~\cite{rottoli2024entanglementhamiltoniansquasiparticlepicture} with a modified occupation function.
Summing over all pairs appearing, we obtain
\begin{equation}\label{eq:quantumEH1}
    K_{\rm q}(t) = \int  \frac{dk}{2\pi} \int dx \chi_{A\overline{A}}(k,x,t) \log\left(\frac{1-\tilde n_\gamma}{\tilde n_\gamma}\right) \hat{n}_{x,k},
\end{equation}
where the counting function $\chi_{A\overline{A}}(k,x,t)$ selects the pairs which contribute, and is identical to the one discussed in Ref.~\cite{rottoli2024entanglementhamiltoniansquasiparticlepicture}.
From \cref{eq:quantumEH1}, we can find the contribution of this term of the entanglement Hamiltonian to the R\'enyi entropies
\begin{equation}
    S_{\rm q}(t) = \int \frac{dk}{2\pi} \min(2|\varepsilon'_k|t,\ell_A) \log(\tilde n_\gamma^\alpha + (1-\tilde n_\gamma)^\alpha)
    \label{quantumentropy}
\end{equation}
which matches exactly the result of Ref.~\cite{alba_carollo2021} for the quantum contribution to the Rènyi entropies.
We remark that in the case under study, the total entanglement entropy is the sum of the entropies deriving from each of the two terms in the entanglement Hamiltonian.
While this is not a generic feature of the sum of entanglement Hamiltonians, it holds in our case because the two terms act on different portions of the effective Hilbert space.

In real space, the solution is again of the form~\eqref{eq:realspaceunitaryK}, where the kernel is modified to a new kernel $\tilde{\mathcal{K}}$ in which the function $\eta(k)$ is replaced by $\tilde{\eta}(k) = \log\left(\frac{1-\tilde n_\gamma}{\tilde n_\gamma}\right)$, 
\begin{align}
    K_{\rm q}(t) &=\int dx  \int dz  \tilde{K}(z,t) c_x^\dag c_{x+z} ,\label{eq:quantumEH} \\
    \tilde{\mathcal{K}}(x,z,t) &= \int \frac{dk}{2\pi}\chi_{A\overline{A}}(k,x,t) \log\left(\frac{1-\tilde n_\gamma}{\tilde n_\gamma}\right)e^{ikz}.
\end{align}
Hence, we conclude that the quantum contribution to the entanglement Hamiltonian during dissipative dynamics can be obtained from the unitary solution through a simple replacement of the occupation functions with their dressed counterparts $\tilde{n}_\gamma$.

\subsection{The classical contribution}
\label{sec:classicalcontr}
While the quantum contribution originates from quasiparticle pairs shared between the subsystem $A$ and its complement, the classical contribution describes pairs that are created within $A$ and for which both quasiparticles remain inside the subsystem.
As we discussed, in the unitary case treated in Ref.~\cite{rottoli2024entanglementhamiltoniansquasiparticlepicture} this term of the reduced density matrix was pure.
Due to the dissipative dynamics, instead, this classical term is now a mixed state too and contributes to the entanglement Hamiltonian. 
We therefore see that this classical term is a novel feature of the dissipative case.
Since neither quasiparticle is traced out when taking the partial trace over the complement, the density matrix of the pair remains unchanged. As a result, to compute the entanglement Hamiltonian it is necessary to express the full density matrix  in~\eqref{eq:evolved_symmetrypreserved_dissipation} as an exponential.
In \cref{appA} we detail the complete calculation of the entanglement Hamiltonian.
Defining $K_{x,k}=-\log\rho_{x,k}$, the final result takes the form
\begin{equation}
\label{eq:classicalKgeneral}
    K^{(x,k)}_{\rm c} = -\log \delta + J\left(e^{i\varphi_k(t)}b_{x_t^+}^\dagger b_{x_t^-}+ e^{-i\varphi_k(t)}b_{x_t^-}^\dagger b_{x_t^+}\right) + \mu_+ \hat{n}_x + \mu_-\hat{n}_{x_t^-},
\end{equation}
where the coefficients $J$ and $\mu_\pm$ are completely determined by the occupation functions and the dressing operation~\eqref{eq:tilded}
\begin{equation}
    \begin{cases}
        J = \sqrt{n_k(1-n_k)} \log\frac{\tilde{0}_\gamma(1-\tilde{1}_\gamma)}{\tilde{1}_\gamma (1-\tilde{0}_\gamma)}, \\
        \mu_{\pm} = \frac{1}{2}\log \frac{(1-\tilde{0}_\gamma)(1-\tilde{1}_\gamma)}{\tilde{0}_\gamma\tilde{1}_\gamma}\pm \frac{1}{2}(1-2n_k) \log\frac{\tilde{1}_\gamma(1-\tilde{0}_\gamma)}{\tilde{0}_\gamma (1-\tilde{1}_\gamma)}.
    \end{cases}
\end{equation}
As expected, the entanglement Hamiltonian~\eqref{eq:classicalKgeneral} is quadratic, preserving the Gaussianity of the state under the Lindbladian evolution. 
While this feature was expected, we note that the calculation reveals it to stem from a very specific fine-tuning of the coefficients, $\alpha \delta=\beta\gamma-\varepsilon^2$, which causes the non-Gaussian contribution to vanish.
In general then $K_{\rm c}^{(x,k)}$ is made of two parts, one which is diagonal in the quasiparticle basis $K_{\rm c}^{\rm d} = \mu_+ \hat{n}_x + \mu_-\hat{n}_{x_t^-}$, and the other which is non-diagonal $K_{\rm c}^{\rm nd}=J\left(b_{x_t^+}^\dagger b_{x_t^-}+b_{x_t^-}^\dagger b_{x_t^+}\right)$.
This off-diagonal term represents an additional difference with respect to the unitary case studied in Ref.~\cite{rottoli2024entanglementhamiltoniansquasiparticlepicture}, where the entanglement Hamiltonian was completely diagonal in the quasiparticle operators.

When considering the Lindbladian evolution of a Néel initial state, the classical part $K^{(x,k)}_{\rm c}$ simplifies considerably, giving 
\begin{equation}\label{eq:kc_neel_simple}\begin{split}
   K^{(x,k)}_{\rm c} =&\, - \log \delta + \frac{1}{2} \log\frac{\tilde{1}_\gamma}{1-\tilde{1}_\gamma} \left (\hat{n}_{x_t^+} + \hat{n}_{x_t^-} + e^{i\varphi_k(t)}b_{x_t^+}^\dagger b_{x_t^-}+e^{-i\varphi_k(t)}b_{x_t^-}^\dagger b_{x_t^+} \right ) \\ &+ \frac{1}{2} \log \frac{\tilde{0}_\gamma}{1-\tilde{0}_\gamma}\left (\hat{n}_{x_t^+} + \hat{n}_{x_t^-}-e^{i\varphi_k(t)} b_{x_t^+}^\dagger b_{x_t^-}-e^{-i\varphi_k(t)}b_{x_t^-}^\dagger b_{x_t^+} \right ). 
\end{split}\end{equation}
{The form in \cref{eq:kc_neel_simple}} is particularly interesting since the terms $\log\frac{\tilde{1}_\gamma}{1-\tilde{1}_\gamma} $ and $\log \frac{\tilde{0}_\gamma}{1-\tilde{0}_\gamma}$ are precisely the functions~\eqref{eq:eta} applied to the occupation functions $\tilde{1}_\gamma$ and $\tilde{0}_\gamma$.
This structure allows us to immediately take the trace in the two-particle subspace, since the two terms constituting $ K^{(x,k)}_{\rm c}$ commute. In particular the entropy contribution of each pair is
\begin{equation}
  \left ( 1-\alpha \right ) S_\text{c}^{\alpha}(x,k) = \log\!\left [\tilde{1}_\gamma^\alpha + \left (1-\tilde{1}_\gamma \right )^\alpha \right ] + \log\!\left [\tilde{0}_\gamma^\alpha + \left (1-\tilde{0}_\gamma\right )^\alpha\right ].
    \label{eq:paircontributionclassical}
\end{equation}

To obtain the total contribution to the entropy of the classical part, consider that the total number of shared pairs at given momentum $k$ is $\frac{1}{2} \max(\ell_A- 2|\varepsilon'_k|t,0)$, leading to 
\begin{equation}
    S^\alpha_{\rm c}(t) = \int_+ \frac{dk}{2\pi}  \max(\ell_A- 2|\varepsilon'_k|t,0) S^\alpha_c(x,k).
    \label{eq:classicalentropy}
\end{equation}
The contribution~\eqref{eq:classicalentropy} to the entanglement entropy corresponds precisely the result of Ref.~\cite{alba_carollo2021} up to a 
different normalization of the energy spectrum. 
If the initial state is not of Néel form, the discussion is slightly more complicated, as to evaluate traces one has to diagonalize~\eqref{eq:classicalKgeneral} in the subspace of a single occupied quasiparticle $\text{span}(\ket{1_{x_t^+},0_{x_t^-}},\ket{0_{x_t^+},1_{x_t^-}}
)$. The eigenvalues are 
\begin{equation}\begin{split}
    \lambda_{\pm} &= \frac{\mu_+ + \mu_-}{2} \pm \sqrt{\left(\frac{\mu_+ - \mu_-}{2}\right)^2 + J^2} \\
    &= \frac{1}{2}\log \frac{(1-\tilde{0}_\gamma)(1-\tilde{1}_\gamma)}{\tilde{0}_\gamma\tilde{1}_\gamma} \pm  \frac{1}{2}\log\frac{\tilde{0}_\gamma(1-\tilde{1}_\gamma)}{\tilde{1}_\gamma (1-\tilde{0}_\gamma)}. 
\end{split}\end{equation}
Note that these are precisely $\log\frac{\tilde{1}_\gamma}{1-\tilde{1}_\gamma} $ and $\log \frac{\tilde{0}_\gamma}{1-\tilde{0}_\gamma}$ again, implying that the entropy contribution of a pair is still of the form~\eqref{eq:paircontributionclassical}.

The total entanglement entropy is then obtained by summing the quantum and classical contributions
\beq
    S^{\gamma}_A(t) = S_{\rm q}(t) + S_{\rm c}(t) 
    \label{eq:entropyqp}
\eeq
where both terms take the same functional form as in Eqs.~\eqref{quantumentropy} and~\eqref{eq:classicalentropy} for all symmetric initial states. 
This result is in agreement with previous findings~\cite{alba_carollo2021,carollo_alba2022}, where the evolution of the entanglement entropy in the presence of dissipation was derived using the correlation matrix techniques.
In the present context, the peculiar structure of the classical contribution to the entropy admits an interesting interpretation in terms of the diagonalization of $K_{\rm c}^{(x,k)}$ in the single-quasiparticle subspace.
The fact that the eigenvalues are $\tilde{0}_\gamma$ and $\tilde{1}_\gamma$ shows that it is possible to perform a change of basis to a set of new degrees of freedom with these occupation functions, which act as normal modes for the dynamics of the classical component.
While in the unitary case such normal modes would be the quasiparticles themselves, in the dissipative case they are expressed as a non-trivial combination of the two as a consequence of the mixing effects.

\begin{figure}[h]
    \centering
    \includegraphics[width=\linewidth]{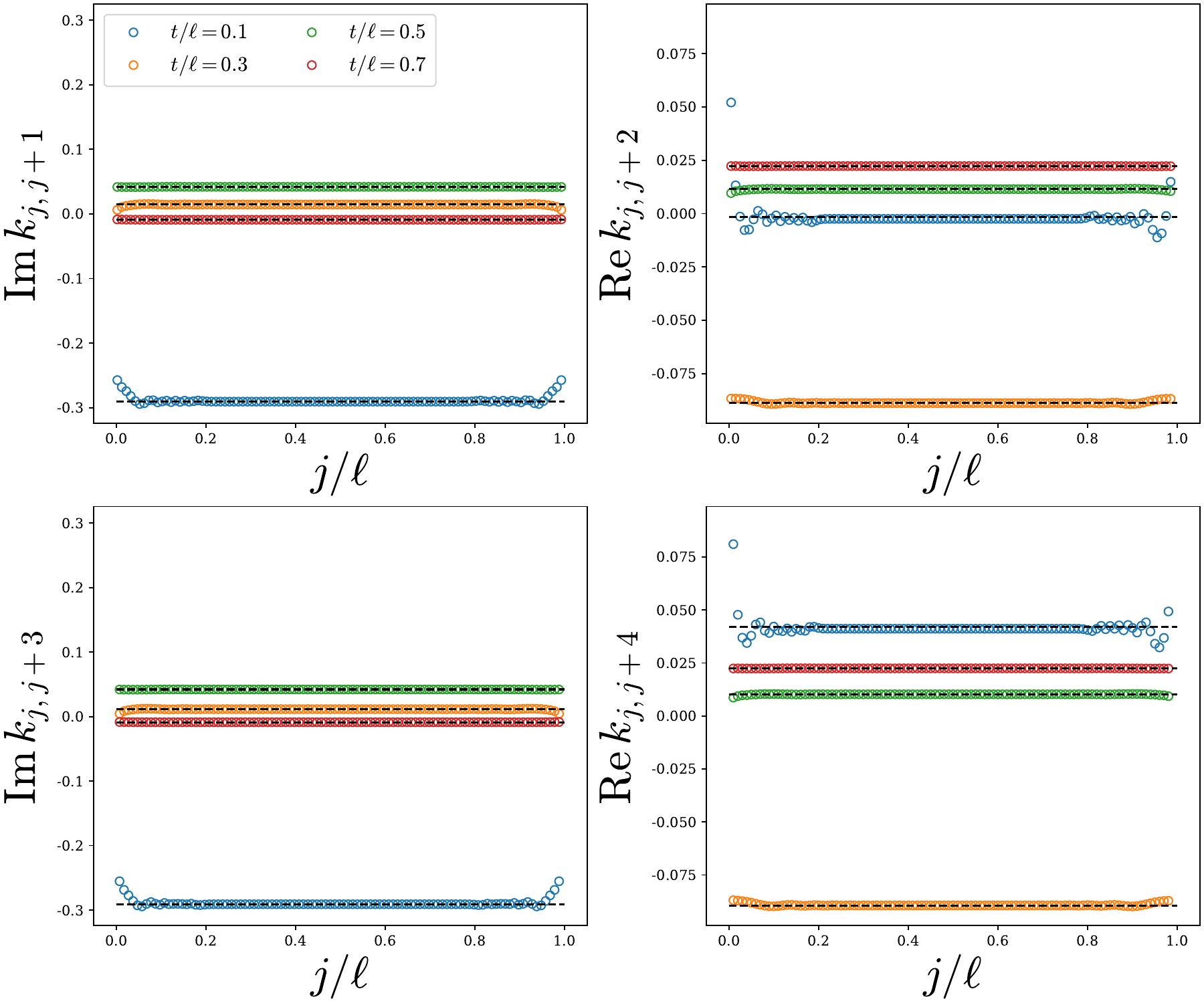}
    \caption{Matrix elements $k_{j,j+z}$ of the entanglement Hamiltonian, with $z$ ranging between 1 and 4 and $j$ restricted to even values to remove the factor $(-1)^j$. The subsystem size is set to $\ell_A=200$, while the dissipation parameters are $\Gamma=0.01$  and $\overline{n}=0.5$ in order to focus on the constant (in $j$) terms. Symbols represent the exact value computed from the correlation matrix, while the dashed lines are the quasiparticle prediction~\eqref{eq:simplified}. The plots show excellent agreement up to small expected deviations at the boundaries.}
    \label{fig:constant}
\end{figure}

Integrating over all modes, the classical part of the EH takes the form 
\begin{equation} \label{eq:kc_generica_qp}
    K_{\rm c}(t) = \int dx\int_{k>0} \frac{dk}{2\pi}\,\chi_{AA}(k,x,t)\, K^{(x,k)}_{\rm c},
\end{equation}
where the counting function selects pairs in which both quasiparticles are contained in $A$
and can therefore be expressed as $\chi_{AA}(k,x,t) = \chi_{[0,\ell_A]}(x) \chi_{[0,\ell_A]}(x-2\varepsilon'_kt)$ where we defined $x$ to be the coordinate of the right mover. 
Note that for $t$ large compared to $\ell_A$, this counting function tends to zero, since asymptotically no quasiparticle pairs remain fully contained within $A$.
Importantly, this behaviour is purely kinematic in origin and therefore independent of the value of $\gamma$, which can be taken arbitrarily small so as to allow the quantum contribution to survive in the long-time limit.
As we will discuss in more detail in the following section, in the weak-dissipation, long-time regime the classical contribution becomes negligible, while the quantum one is saturated from the standpoint of quasiparticle motion, yet retains a residual time dependence induced by dissipation. Before concluding this section, we turn to a real-space reformulation of the result in Eq.~\eqref{eq:kc_generica_qp}.
The diagonal term $ \mu_+ \hat{n}_x + \mu_-\hat{n}_{x_t^-}$ can be dealt with in complete analogy with the result~\eqref{eq:quantumEH}, leading to 
\begin{equation}
    \int dx\int_{k>0} \frac{dk}{2\pi} \chi_{AA}(k,x,t) \mu_+ \hat{n}_{x,k} =  \int dx \int dz \left(\int_{k>0} \frac{dk}{2\pi} \chi_{AA}(k,x,t) \mu_+ e^{ikz}\right) c_x^\dag c_{x+z},
\end{equation}
and similarly for the left moving quasiparticle. 
The off diagonal part is instead more involved, containing a direct coupling of the two modes
\begin{equation}
    K_{\rm c}^{\rm nd} = \int dx \int\frac{dk}{2\pi} \chi_{AA}(k,x,t)J(k)\left( e^{i\varphi_k(t)}b_{x,k}^\dag b_{x-2\varepsilon'_kt,k-\pi} + h.c. \right).
    \label{eq:k_nondiag}
\end{equation}
The presence of the velocity in the fluid cell over which the operators act make the discussion more subtle to be handled analytically; the issue is similar to what appeared for the negativity Hamiltonian considered in~\cite{travaglino2025}. 
The inversion to real space yields\footnote{Here we are being somewhat loose with the distinction between sums and integrals. Within our approximate treatment, this difference is generally negligible, as both the system size and the fluid-cell size are taken to be sufficiently large for sums to be well approximated by integrals. Nevertheless, the appearance of oscillatory factors such as $(-1)^x$ clearly signals underlying lattice effects. }
\begin{equation}
    K_{\rm c}^{\rm nd} = \int dx \,(-1)^x\int dz \int \frac{dk}{2\pi} \chi_{AA}(k,x,t)J(k)e^{i\varphi_k(t)} c_x^\dag c_{x+z -2\varepsilon'_kt}.
\end{equation}
As the kernel exhibits highly oscillatory behaviour for large $t$, the sum over $k$ in~\eqref{eq:k_nondiag} can be approximated by stationary phase arguments, after the identification $z=r-2\varepsilon'_kt$. 
This leads to the approximate value
\begin{equation}
    K_{\rm c}^{\rm nd} \approx \,\sum_x \sum_z (-1)^x i^z \frac{J(k^*)}{\sqrt{\pi t}} \cos\left(2t-z    \frac{\pi}{2}-\frac{\pi}{4}\right)  c_x^\dag c_{x+z}
    \label{eq:simplified}
\end{equation}
where $k^*\approx\frac{z}{2t}$ for $t\gg z$. Note that this is a constant contribution up to the oscillating factors in $z$ and $x$;  this behaviour is indeed observed in the numerical evaluation of figure~\ref{fig:constant}. It is also useful to characterize this correction for the case of Néel and dimer initial states, which will be considered in the following. For the Néel state, the coefficient $J$ has no momentum dependence and therefore the contribution is an oscillating constant proportional to $J_{\rm Neel}=\frac{1}{2}\log\frac{\tilde{0}_\gamma(1-\tilde{1}_\gamma)}{\tilde{1}_\gamma (1-\tilde{0}_\gamma)}$. For dimer states, since $n_k=(1-\cos k)/2$, the parameter is $J(k^*) \sim \frac{z^2}{4t^2}J_{\rm Neel}$; it is therefore clear that in this case the nondiagonal component is strongly suppressed. Both of these predictions are confirmed numerically in figures~\ref{fig:constant},~\ref{fig:comparison1} and~\ref{fig:dimercomparison}.

\section{Time-dependent GGE}
\label{sec:tGGE}
Combining the two contributions evaluated in the previous section, we finally obtain the expression for the entanglement Hamiltonian in the coarse-grained basis
\begin{equation}\label{eq:kfull}
   K_A(t)=  K_{\rm q}(t) + K_{\rm c}(t).
\end{equation}
As anticipated, in this section we study the asymptotic behaviour of the entanglement Hamiltonian~\eqref{eq:kfull} at long times $t \gg \ell_A$, taking the scaling limit of weak dissipation $\Gamma t= \text{const}$.
Asymptotically, at long times all quasiparticle pairs escape the subsystem, and as a consequence the Hilbert space associated with pairs fully contained in $A$ shrinks to zero dimension, causing the classical component of the entanglement Hamiltonian~\eqref{eq:kfull} to vanish identically.
On the other hand, in this regime the quantum part yields
\begin{equation}
    K_A(t\gg \ell_A)=  K_{\rm q}(t\gg \ell_A) = \lim_{\overset{t\to \infty}{\Gamma t=\text{const}}}\int  \frac{dk}{2\pi} \int dx\, \chi_{A\overline{A}}(k,x,t)\, \log\!\left(\frac{1-\tilde n_\gamma}{\tilde n_\gamma}\right) \hat{n}_{x,k}. 
\end{equation}
Following the discussion of Ref.~\cite{rottoli2024entanglementhamiltoniansquasiparticlepicture}, the limit is most conveniently obtained by reverting to the real space fermion representation, which results in 
\begin{equation}
     K_A(t\gg \ell_A)= \int  \frac{dk}{2\pi}\,  \log\!\left(\frac{1-\tilde n_\gamma}{\tilde n_\gamma}\right) c_k^\dag c_k\label{eq:ehtGGE},
\end{equation}
which is precisely a time-dependent GGE (t-GGE), where the time dependence enters exclusively through the evolution of the occupation number $\tilde n_\gamma= e^{-\Gamma t}n_k +\overline{n} (1-e^{-\Gamma t})$.
Let us comment on the significance of this result.
\cref{eq:ehtGGE} shows that the quasiparticle picture naturally gives rise to time-dependent GGEs in weakly dissipative fermionic systems and explains the origin of this effect.
By taking the limit  $t \to \infty$ with $\Gamma t= \text{const}$, as is standard in the literature on the subject, the contribution $K_{\rm c}(t)$, which contains correlations beyond a GGE structure, goes to zero leaving only the quantum component, which indeed reduces to a t-GGE. 
We stress that the emergence of the t-GGE only requires the time to be large compared to the subsystem size $\ell_A$, which is the same regime of emergence of the standard GGE in unitary systems. 
Our framework clearly demonstrates that the dynamics separates into a fast unitary evolution, which drives relaxation toward a GGE, and a slow dissipation-induced evolution between different GGEs, parameterized by the occupation functions $\tilde{n}_\gamma$. 

We emphasize that, within this framework, the origin of the t-GGE can be traced back to the simple structure of the dissipative term in the Lindbladian when expressed in terms of the modes $b_{x,k}$, which constitute the most natural degrees of freedom for describing quench dynamics at this scale. 
Together with the assumption of Gaussianity, required to evaluate the correlators, this is the only property of the evolution that enters our derivation. It is therefore expected that any dynamics sharing these features would lead to conclusions analogous to those obtained in this work. 

\section{Numerical evaluations}
\label{sec:numerics}
In this section, we numerically investigate an example of quenches with dissipative dynamics and compare the results with the theoretical predictions. 
Since the gain and loss dissipation acts on the correlation matrix simply as
\begin{equation}
    C_{xy}(t) =  e^{-\Gamma t} C_{xy}^{\rm u}(t) + \overline{n}(1-e^{-\Gamma t}) \delta_{x,y},
\end{equation}
the numerical approach follows directly the discussion of Refs.~\cite{rottoli2024entanglementhamiltoniansquasiparticlepicture,travaglino2024quasiparticlepictureentanglementhamiltonians} with a simple modification. 
In particular, the entanglement Hamiltonian can be expressed in terms of the correlation matrix as
\begin{equation}
     K_A(t) = \log \frac{1-C_A(t)}{C_A(t)},
\end{equation}
where $C_A(t)$ is the restriction of the correlation matrix to the subsystem of interest.
Since the inclusion of dissipation shifts the eigenvalues of the correlation matrix toward $\overline{n}$, there is no need to introduce the cutoff employed in~\cite{rottoli2024entanglementhamiltoniansquasiparticlepicture} to handle the pure component of the density matrix. 

\begin{figure}[t!]
    \centering
    \includegraphics[width=\linewidth]{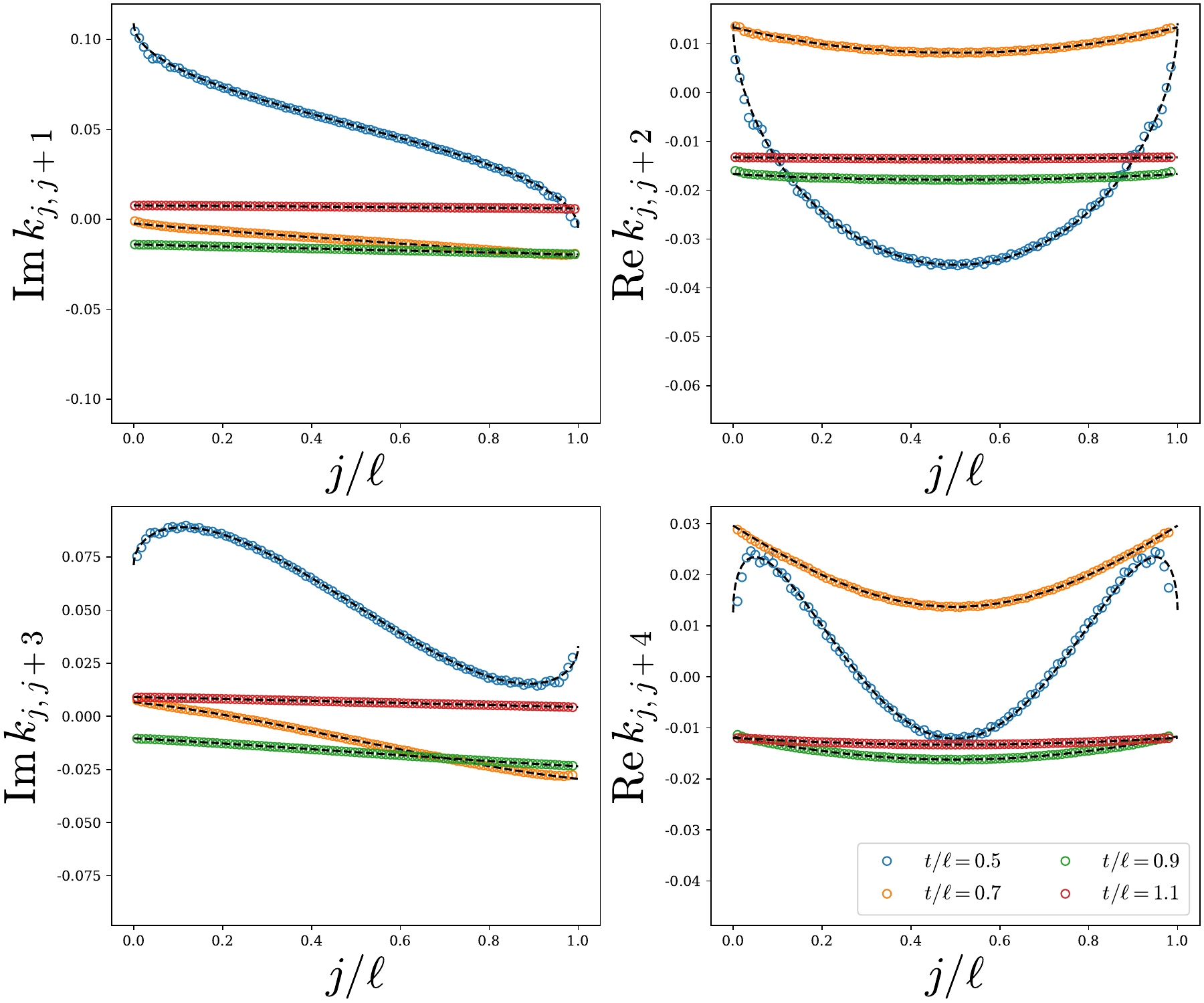}
    \caption{EH in a quench from the Néel state on a lattice of $\ell_A = 200$ sites, with dissipation parameters $\Gamma = 0.01$ and $\overline{n}=0.8$. In order to avoid the oscillating factors, only even values of x are shown. It is clear that for $\overline{n} \neq 0.5$ the solution is no longer constant in $x$, but the extra terms are fully accounted for by the addition of $K_{\rm q}(t)$ and $K_{\rm c}^{\rm d}$.}
    \label{fig:comparison1}
\end{figure}

\begin{figure}[t!]
    \centering
    \includegraphics[width=\linewidth]{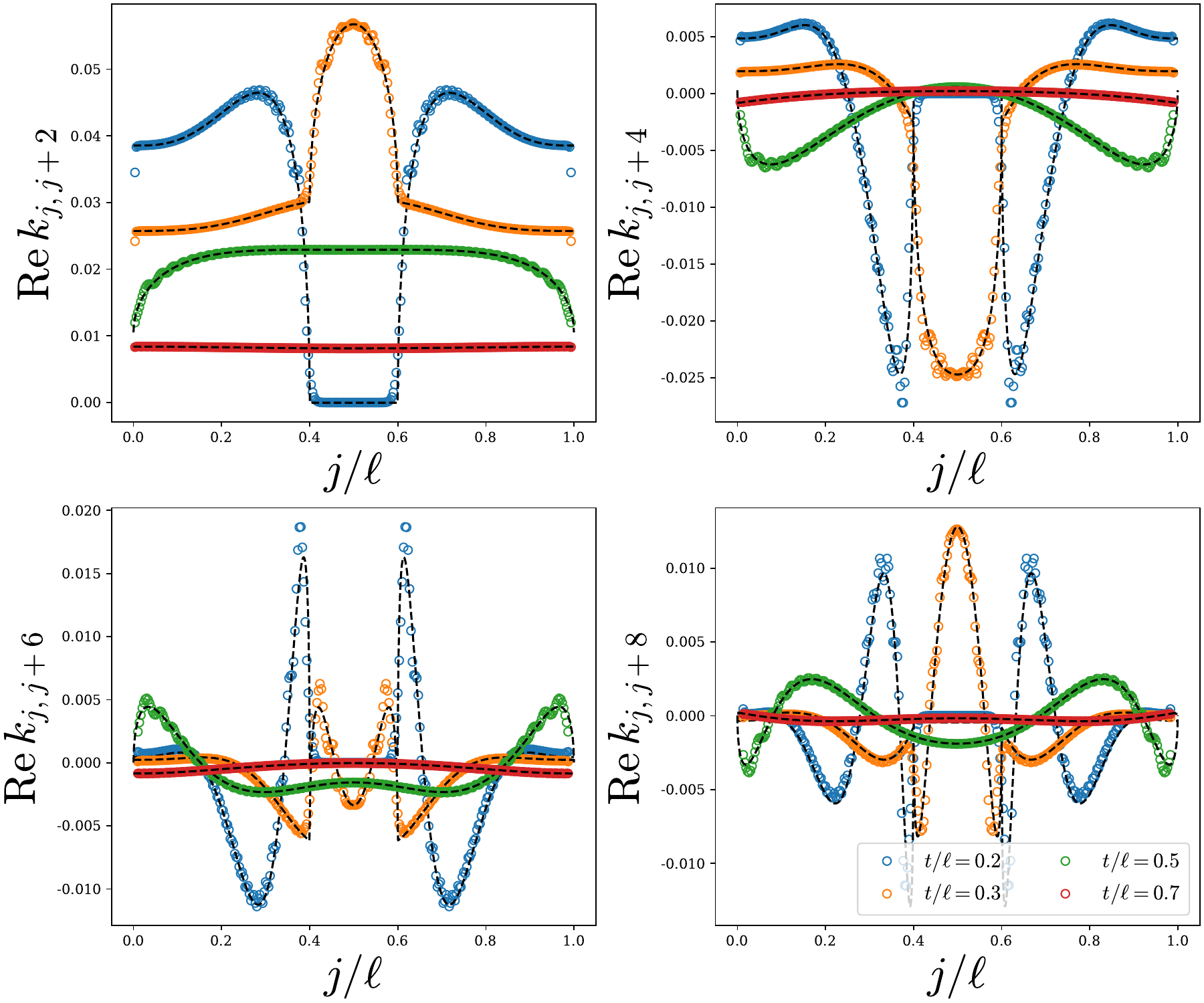}
    \caption{Elements of the EH for a quench from a dimer state, with $\Gamma = 0.01$ and $\overline{n}=0.2$, for subsystem size $\ell_A=300$. The dashed analytical curve is obtained simply by considering the quantum part and the diagonal component of the classical part; the agreement shows that the non-diagonal part is strongly suppressed for this initial state, as argued in section~\ref{sec:classicalcontr}. }
    \label{fig:dimercomparison}
\end{figure}

We consider quenches from the Néel state $\ket{0101\dots}$, for which the theoretical prediction is greatly simplified, as $\mu_+=\mu_-$. 
The unitary correlation matrix is, in the limit of large total system size,
\begin{equation}
    C_{xy}(t) = \frac{1}{2}\left(\delta_{x,y} + i^{x-y}(-1)^yJ_{x-y}(2t)\right),
\end{equation}
where we have introduced the Bessel function of first kind $J_x(z)$. In this case the classical contribution $K_{\rm c}(t)$ highly simplifies as shown above: the coefficients $\mu_{\pm}$ and $J$ lose all momentum dependence and reduce to simple functions of $\tilde{0}_\gamma$ and $\tilde{1}_\gamma$, and the off-diagonal part can be simplified as shown in~\eqref{eq:simplified}. 
We begin by isolating and testing this specific contribution. 
This can be achieved by setting $\overline{n}=1/2$ so that both the quantum part and the diagonal component of the classical part vanish (the first because $\tilde{n}_\gamma=1/2$, the second because $\mu_{\pm}=0$ when $\overline{n}=1/2$). 
Hence, in this regime, only $K_{\rm c}^{\rm nd}$ survives in the entanglement Hamiltonian, as shown in \cref{fig:constant}. 
In order to avoid to deal with the factor $(-1)^x$ and to focus on the constant value, we plot only even values of  $x$. 
The numerical results show clearly that the value~\eqref{eq:simplified} describes correctly the constant value of the correlators. 
For $\overline{n} \neq 0.5$, the other two terms become relevant, but $K_{\rm c}^{\rm nd}$ still contributes as an oscillating constant. Focusing again on a fixed parity of $x$, the result for a given choice of parameters are shown in \cref{fig:comparison1}, highlighting an excellent agreement between the analytical prediction and the exact solution up to small corrections at the boundaries.

\begin{figure}[t!]
    \centering
    \includegraphics[width=\linewidth]{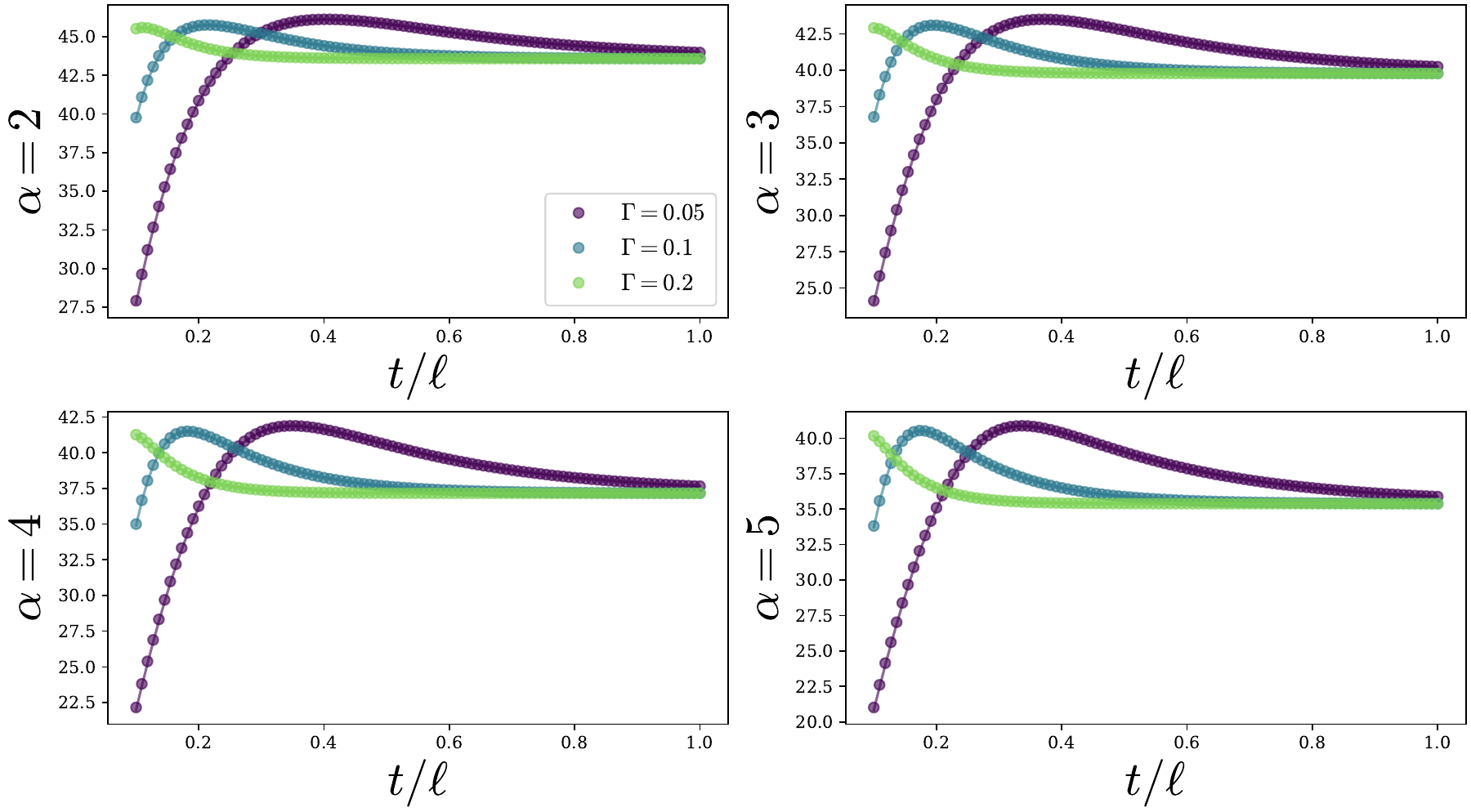}
    \caption{Renyi entropies under dissipative dynamics after a quench from the dimer state, for several values of $\Gamma$ at fixed $\overline{n}=0.3$, for a subsystem of $\ell_A=80$ sites. 
    The symbols represent the exact results obtained through~\eqref{eq:renyicorr}, while the lines are the quasiparticle prediction~\eqref{eq:entropyqp}. The plot shows a perfect agreement even for a relatively small subsystem size. Note the saturation to a $\Gamma$-independent entropy with fixed occupation function $\overline{n}$.  }
    \label{fig:renyi_dimer}
\end{figure}

It is also interesting to consider a quench from the dimer state
\begin{equation}
\ket{\psi} \propto \prod_{i \text{ even}} \left(\ket{1_i0_{i+1}} \pm \ket{0_i 1_{i+1}} \right),
\end{equation} 
in which nearest neighbours are entangled in Bell pairs. Under unitary evolution governed by the hopping Hamiltonian, the correlation matrix  evolves as~\cite{Eisler_2007}
\begin{equation}
    C^{\rm u}_{x,y}(t) =C_{x,y}^{(\infty)} + i \frac{x-y}{4t}e^{-i\frac{\pi}{2}(x+y)} J_{x-y}(2t),
    \end{equation}
where $C_{x,y}^{(\infty)} = \frac{1}{2}\delta_{x,y} + \frac{1}{4}(\delta_{x,y+1}+\delta_{x,y-1})$, corresponding to the occupation functions $n_k= (1-\cos k )/2$. As discussed in section~\ref{sec:classicalcontr}, in this situation the term $K_{\rm c}^{\rm nd}$ is approximately zero, and therefore the solution is fully determined by the sum of $K_{\rm q}$ and $K_{\rm c}^{\rm d}$. 
In \cref{fig:dimercomparison} the exact result is compared to the analytical prediction for $K_{\rm q}+K_{\rm c}^{\rm d}$ showing a perfect agreement. 
In particular, the corrections given by $K_{\rm c}^{\rm nd}$ are of order $t^{-5/2}$ and are completely invisible at this scale.
We stress that the restriction to small $\Gamma$ in the plots is purely due to numerical considerations, since the analytical results hold for arbitrary $\Gamma$. In practice, accessing the ballistic regime requires large $\ell_A$ and $t$, so $\Gamma$ must be chosen small only to avoid numerical instabilities in evaluating exponentials.

Correlation matrix techniques also allow us to obtain the Rényi entropies directly. In fact, these are determined by the spectrum of the correlation matrix $\{\lambda_j\}$ as 
\begin{equation}
    S_A^{(\alpha)} = \frac{1}{1-\alpha} \sum_i \log(\lambda_i^\alpha + (1-\lambda_i)^\alpha).
    \label{eq:renyicorr}
\end{equation}
Given the solution for the correlation matrix, at long times entanglement is expected to saturate to a value fixed by $\overline{n}= \gamma_G /\Gamma$, as shown in \cref{fig:renyi_dimer} reporting several values of $\Gamma$ at fixed $\overline{n}$. In the plot, circles represent the exact solution obtained from~\eqref{eq:renyicorr}, while the lines are the analytical prediction~\eqref{eq:entropyqp}. The curve show the perfect agreement for several values of Rènyi index and of dissipation, thus confirming the results of~\cite{carollo_alba2022}.

\section{Conclusions}
\label{sec:concl}
In this work, we have analysed the post-quench dynamics of free fermionic systems subject to gain-and-loss dissipation. By reformulating the Lindblad evolution in terms of coarse-grained fermionic degrees of freedom, we solved the dynamics at leading order in the ballistic regime and derived an explicit expression for the entanglement Hamiltonian, recovering previously known results for Rényi entropies.
We showed that the entanglement Hamiltonian naturally decomposes into two distinct contributions: a purely classical component, entirely induced by dissipation, and a quantum component originating from unitary correlations dressed by dissipative effects. In the long-time limit, the classical contribution vanishes as a consequence of quasiparticle motion, while the quantum component approaches a time-dependent generalized Gibbs ensemble (t-GGE) under an appropriate scaling of the dissipation strength. The emergence of the t-GGE is thus directly traced back to the quasiparticle picture, and our results also characterize the intermediate time scales governing the relaxation toward this regime.

Several interesting directions remain open. One natural extension concerns systems undergoing local measurements, where the quasiparticle picture is known to break down at the level of Rényi entropies~\cite{carollo2022}. In such cases, this breakdown may be reflected in a more intricate correlation structure among the operators $b_{x,k}$ and $b_{x,k}^\dag$, as well as in a more complex effective propagation. Another promising direction is to combine the present results with those of Ref.~\cite{travaglino2025} to gain insight into the negativity Hamiltonian. This would be particularly relevant, as negativity-based measures are more sensitive probes of entanglement in mixed states than standard Rényi entropies.
A final interesting open question is whether some of the qualitative features of the entanglement Hamiltonian identified here persist in systems that are not analytically tractable using Gaussian techniques, such as those with more complex forms of dissipation or with interactions in the Hamiltonian dynamics.

\section*{Acknowledgement}
We thank Viktor Eisler and Vincenzo Alba for useful discussions on related topics.
P.C. and R.T. acknowledge support from the European Research Council under the Advanced Grant no. 101199196 (MOSE). 
F.R. acknowledges support from the project “Artificially devised many-body quantum dynamics in low dimensions - ManyQLowD” funded by the MIUR Progetti di Ricerca di Rilevante Interesse Nazionale (PRIN) Bando 2022 - grant 2022R35ZBF.

\appendix
\section{Derivation of the ballistic evolution}
\label{App}
As discussed in the main text, the core of our approach relies on introducing coarse-grained degrees of freedom $b_{x,k}$ that evolve ballistically at leading order, giving rise to the quasiparticle picture, while exhibiting dispersive corrections at higher orders. In this section, we detail how this approximation is constructed and we derive \cref{eq:QPP_heisenber_evolution}. 
The evolution of $b_{x,k}$ is simply given by
\begin{align}
    b_{x,k}^\dag (t) &=\frac{1}{\sqrt{\Delta}} \sum_z e^{-ikz} c_{x+z}^\dag(t) = \frac{1}{\sqrt{\Delta L}}\sum_{z} \sum_q  e^{-ikz} e^{iq(x+z)} e^{i\varepsilon_qt} c_q^\dag \\
    &=\frac{1}{L \sqrt{\Delta}}  \sum_{z} \sum_q \sum_y e^{-ikz} e^{iq(x+z)} e^{i\varepsilon_qt} e^{-iqy} c_y^\dag.
\end{align}
Importantly, the structure of the packet implies that the momenta $q$ that contribute significantly in the above expression are those close to $k$. This fact allows us to set up a series expansion in $q-k$, and in particular to write $\varepsilon_q \approx \varepsilon_k +(q-k) \varepsilon_k' + \frac{(q-k)^2}{2} \varepsilon_k'' + \dots$. The first term in the expansion corresponds to the situation in which $\Delta$ consists of the full system, and therefore the $b_{x,k}$ reduce to the standard momentum space fermions. The first nontrivial order gives the ballistic evolution, while the following order gives the leading dispersive contributions. Considering up to first order, one gets
\begin{align}
    b_{x,k}^\dag (t)  &\approx \frac{e^{it\varepsilon_k}}{L \sqrt{\Delta}}  \sum_{z,y} e^{-ikz} \sum_q\left(   e^{iq(x+z)} e^{i(q-k)\varepsilon'_kt} e^{-iqy} \right) c_y^\dag \\
    &= 
  \frac{1}{\sqrt{\Delta}}  e^{it\varepsilon_k} \sum_z e^{-ikz} c^\dag_{x+z + \varepsilon'_kt} = e^{it\varepsilon_k} b_{x+\varepsilon'_kt,k}^\dag ,
\end{align}
where we have used that the sum over $q$ yields a delta function, and that $e^{ikX} = 1$ whenever $X$ is a fluid cell coordinate. 
Higher orders in the expansion lead to dispersive effects. These give rise to sublinear contributions to the EH, and therefore can be neglected for our purposes. 
Similar conclusions can be reached by considering Gaussian wave packets instead of the square packets used here. This choice leads to an entirely analogous evolution, provided that the width of the Gaussian is large compared to microscopic length scales but still small relative to macroscopic ones. This demonstrates that different coarse-graining procedures can be adopted, depending on the physical context, while yielding identical results.

\section{\texorpdfstring{Exponentiation of $K_{\rm c}(t)$}{Exponentiation of Kc(t)}}
\label{appA}
In this appendix we provide some more details on the construction of the EH for the classical part, obtained by pairs in which both quasiparticles are in the subsystem $A$ at the time $t$. The starting point is the density matrix~\eqref{eq:evolved_symmetrypreserved_dissipation}, 
\begin{align}
    \rho =&\, \bigotimes_{x,k} \rho_{x,k}(t), \\
    \rho_{x,k}{(t)} =&\,  \alpha \hat{n}_{x_t^+}\hat{n}_{x_t^-} + \beta \hat{n}_{x_t^+}(1-\hat{n}_{x_t^-} )+\gamma (1-\hat{n}_{x_t^+} ) \hat{n}_{x_t^-}\nonumber\\
    &+  \delta (1-\hat{n}_{x_t^+} )(1-\hat{n}_{x_t^-} )  + \varepsilon\left(e^{i\varphi_k}b_{x_t^+}^\dagger b_{x_t^-}+e^{-i\varphi_k}b_{x_t^-}^\dagger b_{x_t^+}\right),    
\end{align}
where the coefficients are presented in \cref{eq:coefficients}. We are interested in finding the operator whose exponential reproduces $\rho_{x,k}(t)$, which would represent the contribution to the EH of a single pair. This can be done by considering $\rho_{x,k}$ as an operator acting on the Hilbert space of the two quasiparticles, spanned by $\ket{i,j}$ where $i,j\in\{0,1\}$ are the occupations. In this space, $\rho_{x,k}$ can be represented as a matrix
\begin{equation}
    (\bra{i,j}\rho_{x,k}\ket{k,l}) = \begin{pmatrix}
        \delta & 0 &0&0\\
        0& \beta & e^{i\varphi_k(t)}\varepsilon &0\\
        0&  e^{-i\varphi_k(t)}\varepsilon &\gamma &0\\
        0&0&0& \alpha
    \end{pmatrix},
\end{equation}
in which the subspaces with $0$ and $2$ particles can be written in exponential form directly, while the $1$ particle sector needs to be diagonalized first. The eigenvalues in this sector are $ \theta_{\pm}= \frac{\beta + \gamma}{2} \pm \sqrt{(\frac{\beta - \gamma}{2})^2 + \varepsilon^2}$, corresponding to eigenvectors $\ket{\psi_{\pm}}=  e^{i\varphi_k(t)/2}\sqrt{n} \ket{10} \pm  e^{-i\varphi_k(t)/2}\sqrt{1-n}\ket{01}$, and therefore the logarithm of $\rho_{x,k}$ is
\begin{equation}
    \log \rho_{x,k} = \log \delta \ket{00}\bra{00} + \log \theta_+ \ket{\psi_+} \bra{\psi_+ } +  \log \theta_- \ket{\psi_-} \bra{\psi_- } + \log \alpha \ket{11}\bra{11}.
\end{equation}
To connect with the result of the main text, it is enough to notice that $\theta_+ = \tilde{1}_\gamma (1-\tilde{0}_\gamma)$ and  $\theta_- = \tilde{0}_\gamma (1-\tilde{1}_\gamma)$, and to use $\ket{00}\bra{00}  = (1-n_{x_t^+})(1-n_{x_t^-})$, $\ket{11}\bra{11}= n_{x_t^+} n_{x_t^-}$, and similarly for the other terms. Performing such replacements one obtains
\begin{equation}
     \log \rho_{x,k}  = \log\delta - \mu_+ \hat{n}_{x_t^+ }- \mu_- \hat{n}_{x_t^- } - J (e^{i\varphi_k(t)}b_{x_t^+} b_{x_t^-}+ h.c.) + \log\left(\frac{\alpha \delta}{\beta \gamma - \varepsilon^2}\right) n_{x_t^+} n_{x_t^-}.
\end{equation}
Upon noticing that $\alpha \delta = \beta \gamma - \varepsilon^2$, the non-Gaussian component vanishes, leading to the claimed result for $K_{\rm c}^{(x,k)} = - \log \rho_{x,k}$.

\section{Symmetry breaking states}
\label{appB}
In this appendix, we consider the generalisation to symmetry breaking initial states, showing that the main arguments are identical.
For simplicity, we focus on pure loss dissipation, as the generalization is straightforward.
All the main arguments of the derivation in the main text also apply to this context, with the caveat that the initial state is of the form~\eqref{eq:coherent}.
The occupation functions immediately evolve as
\begin{equation}
     \braket{\hat{n}_{x,k}(t)} = e^{-\gamma t}  \braket{\hat{n}_{x,k}(t)}_{\rm unitary} = e^{-\gamma t} n_k.
\end{equation}
In a similar way, we can evaluate the evolution of products of number operators at different sites, using Wick theorem. Regarding the interesting ones, corresponding to right and left moving pairs, we obtain
\begin{equation}
    \braket{\hat n_{x_t^+} \hat n_{x_t^-}} = n_k e^{-2\gamma t}(2n_k-1) 
\end{equation}
from which it is then possible to get the cross terms $\braket{n_{x_t^+} (1-n_{x_t^-})} = \braket{(1-n_{x_t^+}) n_{x_t^-}} = ne^{-\gamma t} (1-e^{-\gamma t}(2n-1))$ and finally $\braket{(1-n_{x_t^+})(1-n_{x_t^-})} = 1-2ne^{-\gamma t} + ne^{-2\gamma t}(2n-1)$.
At this point, with the same arguments as in the main text, the density matrix is again given by
\begin{equation}
    \rho(t) = \bigotimes_{x,k} \rho_{x,k}(t),
\end{equation}
where $\rho_{x,k}$ is in the form~\eqref{eq:evolved_symmetrypreserved_dissipation}, with different coefficients which can be determined by the new values of the correlators. 

Putting the pieces together, the density matrix of a pair in the dissipative regime is
\begin{equation}\begin{split}
    \rho_{x,k} =&\, n_k e^{-2\gamma t}  (2n_k-1)\hat{n}_{x_t^+}\hat{n}_{x_t^-} + e^{-\gamma t}\sqrt{n(1-n)}( b_{x_t^+}^\dagger b_{x_t^-} ^\dagger + h.c.) \\ 
    &+ n_k e^{-\gamma t}(1-e^{-\gamma t}(2n_k-1))\left[(1-\hat{n}_{x_t^+})\hat{n}_{x_t^-} +  \hat{n}_{x_t^+}(1-\hat{n}_{x_t^-}) \right]  \\
    &+ \left[1- 2n_k e^{-\gamma t}   +    n_k e^{-2\gamma t}(2n_k-1) \right] (1-\hat{n}_{x_t^+})(1-\hat{n}_{x_t^-}) . \label{densitymatrixdissipation}
\end{split}\end{equation}
It is now possible to construct the EH, similarly to what was done above, by considering a quantum and a classical part. These have similar properties to the symmetric case, with the main difference arising from the presence of the term $b^{\dag}_{x_t^+} b^{\dag}_{x_t^-}$ which breaks particle number conservation. As the construction is essentially identical, we will only consider the quantum part to show that the obtained t-GGE is of the same form despite the different structure. For shared pairs, in which the right mover is in $A$ and the left mover in $\overline{A}$, the trace over the complement gives
\beq
    \Tr_{\overline{A}}\left[ \rho^A_{x_0,k}(t)\right] =
 \tilde{n}_\gamma \hat{n}_{x_t^+} + (1- \tilde{n}_\gamma)(1-\hat{n}_{x_t^+}),
 \label{eq:sharedsqueezeddiss}
\eeq
where again we have introduced $\tilde{n}_\gamma$, which in this case is simply $\tilde{n}_\gamma = ne^{-\gamma t}$ because of the pure loss regime. This can then be written in exponential form as above,
\beq
      \Tr_{\overline{A}}\left[ \rho^A_{x_0,k}(t)\right]= (1-\tilde{n}_\gamma \exp\left\{-\eta_\gamma(t,k)\hat{n}_{x,k}\right\}.
\eeq
 Hence the quantum part of the EH is the same as for symmetric states,
\begin{equation}
    K_{\rm q}(t) = \int dx \int \frac{dk}{2\pi} \chi_{A\overline{A}}(x,k,t) \eta_\gamma(t,k)\hat{n}_{x,k},
\end{equation}
which implies that in the limit $t\gg \ell_A$, $\gamma t=\text{const}$ we have the same t-GGE as obtained previously.
Exponentiation of the classical contributions can then be performed on the same lines as for symmetric states, giving as a general solution 
\begin{equation}
    K_{\rm c}^{(x,k)} = J' (b^\dag_{x_t^+}b^\dag_{x_t^-} + h.c.) + \mu'_+ \hat{n}_{x_t^+} + \mu'_- \hat{n}_{x_t^-}.
\end{equation}

\bibliographystyle{ytphys}
\bibliography{bibliography}
\end{document}